\documentclass[11pt]{article} 

\usepackage[latin1]{inputenc}
\usepackage[OT1]{fontenc}
\usepackage[english]{babel}
\usepackage{indentfirst}
\usepackage{aeguill}
\usepackage{amsmath, amssymb}
\usepackage{array}

\begin{document}

\title{A Distributed Trust Diffusion Protocol for Ad Hoc Networks}

\author{
  Michel Morvan$^{1,2,4}$ and Sylvain 
  Sené$^{3,4}$\thanks{Corresponding author. Email address: Sylvain.Sene@ens-lyon.fr}\\\\
  $^1$~Université de Lyon, LIP,\\
  46, allée d'Italie, 69364 Lyon cedex 07, France\\
  $^2$~\'Ecole des hautes études en sciences sociales and Santa Fe 
  Institute\\[0.1cm]
  $^3$~UJF-Grenoble, TIMC-IMAG,\\
  Faculté de médecine, 38706 La Tronche cedex, France\\
  $^4$~IXXI, Institut rhône-alpin des systèmes complexes\\
  5, rue du Vercors, 69007 Lyon, France
}

\date{}

\maketitle
\thispagestyle{empty}

\begin{abstract}
  In this paper, we propose and evaluate a distributed protocol to
  manage trust diffusion in ad hoc networks. In this protocol, each node $i$ 
  maintains a ``trust value'' about an other node $j$ which is computed both as
  a result of the exchanges with node $j$ itself and as a function of the 
  opinion that other nodes have about $j$. These two aspects are respectively 
  weighted by a trust index that measures the trust quality the node has in its
  own experiences and by a trust index representing the trust the node has in 
  the opinions of the other nodes. Simulations have been realized to validate 
  the robustness of this protocol against three kinds of attacks: simple 
  coalitions, Trojan attacks and detonator attacks.
\end{abstract}

\section{Introduction and preliminaries}
\label{intro}

{\em Ad hoc networks} can be defined as collections of mobile nodes, 
distributed and independent, being able to communicate by radio transmission 
and to self-organise. They constitute networks with unstable infrastructure.
Since the neighbourhood of every node changes over time, it is important to 
develop protocols that will help each node to identify reliably other nodes 
with which it can interact safely. 
\medskip 

Let us present two definitions of the concept of {\em trust} given by the 
Webster's classic 1913: 
\textit{``Firm reliance on the integrity, ability, or character of a person or 
  thing''} and \textit{``Certainty based on past experience''}. Trust is 
therefore a natural notion acquired and systematically used by anyone to decide
if an exchange with somebody else is conceivable or not. Let us remark that the
trust we can have in somebody usually depends on the personal knowledges we 
have about the person and also on his/her reputation according to other 
people we have already met. This is usually the way that trust diffuses in 
human groups. For example, a researcher can trust the scientific opinion of a 
colleague while not trusting the opinion of the same colleague about the work 
of other scientists. The idea of this paper is to transfer this human trust 
diffusion protocol to ad hoc networks. Literature about trust management in 
P2P systems~\cite{buchegger,sepandar} and Web Services~\cite{ebay} presents 
some solutions using both 
personal and external opinions. However, these protocols do not offer 
protection against attacks made by a set of nodes working in coalition, for 
example diffusing wrong positive opinions about a dishonest node (called Trojan
attack further in the paper). That is why we propose  to study in this paper a 
management policy that also measures the trust quality by introducing a 
{\em new level of trust} that adds weights both on a node own experiences and 
on external knowledges this node has received from the others. 

Let us call {\em efficient} a protocol that leads to prevent bad interaction 
and {\em reliable} a protocol that aims at favouring good interactions. This 
work consists in creating, studying and validating by simulations a new 
efficient distributed trust diffusion protocol for ad hoc networks using the 
double level of trust described above. 
\smallskip

Section~\ref{model} presents and explains the trust diffusion protocol. Some 
results obtained on this protocol are exposed in Section~\ref{simul}. In 
particular, this section shows how the protocol can resist to three different 
types of attacks. Section~\ref{perspec} gives perspectives of this work.

\section{Trust diffusion protocol}
\label{model}

Because of the constraints inherent to ad hoc networks, the proposed model has 
to be based on three fundamental concepts: information decentralisation, 
diffusion and management of trust histories. 

In this section, we first present the different attributes used by the protocol
in order to have the requested features to adapt on ad hoc networks. Then, we 
expose the protocol dynamics.  

\subsection{Protocol attributes}

In order to respect the constraints induced by ad hoc networks, all nodes have 
to store their own knowledges about the others. This subsection introduces 
the different attributes that compose the notion of {\em knowledge} used in the
proposed protocol and that a node $i$ has about a node $j$ at time $t$.
{\vskip 1.5mm}

\noindent \textbf{$\bullet$ Trust mark}, denoted by $tm_{i,j}(t) \in \lbrack 
0,1 \rbrack$.

Node $i$ stores a trust mark valued in $\lbrack 0,1 \rbrack$ about node $j$ 
with which it has already interacted. This trust mark is used to know if an 
interaction has a good chance to be benefic. It is updated as a function of 
two major information: the own experience of node $i$ and the external 
knowledges node $i$ has obtained from other nodes during past interactions. It 
is initialized to $\frac{1}{2}$.
{\vskip 1.5mm}

\noindent \textbf{$\bullet$ Trust index in trust mark}, denoted by 
$itm_{i,j}(t) \in \lbrack 0,1 \rbrack$.

Node $i$ stores a trust index about its trust mark about node $j$. This trust 
index gives an indication about the reliability of the trust mark it has about 
node $j$. It is initialized to $\frac{1}{2}$.
{\vskip 1.5mm}

\noindent \textbf{$\bullet$ List of external trust knowledges}, denoted by 
$et_{i,j}(t)$ with $et_{i,j}(t)[k] \in \lbrack 0,1 \rbrack$ and $k \in [0,n]$.

Node $i$ stores about node $j$ a list of external trust marks. Such a list
contains the trust marks owned by $j$ about other nodes. This list has been 
transmitted to node $i$ during the last interaction between $i$ and $j$ (during
this interaction, $i$ has symmetrically transmitted its trust marks to $j$). 
This list is initialized to $\emptyset$. Moreover, when nodes $i$ and $j$ 
symmetrically transmit to each other their trust marks, they also transmit their
trust indices in these trust marks. These lists of trust indices are called 
lists of external trust indices in trust knowledges. Such a list owned by node 
$i$ about node $j$ is denoted at time $t$ by $iel_{i,j}(t)$ with 
$iel_{i,j}(t)[k] \in [0,1]$ and $k \in [0,n]$ and is initialized to 
$\emptyset$.  
{\vskip 1.5mm}

\noindent \textbf{$\bullet$ Trust index in list of external trust knowledges}, 
denoted by $iet_{i,j}(t) \in [0,1]$.

Node $i$ stores a trust index about each list of external knowledges it 
has received from other nodes. Such trust indices give an indication about 
the reliability of these lists and are initialized to $\frac{1}{2}$.
{\vskip 1.5mm}

\noindent \textbf{$\bullet$ List of past interactions marks}, denoted by 
$pil_{i,j}(t)$ with $pil_{i,j}(t)[k] \in \lbrace 0,1 \rbrace$ and $k \in 
\lbrack 0,\theta \rbrack$.

Node $i$ stores about node $j$ a list of marks encoding the results of its 
$\theta$ (where $\theta$ is bounded by $\theta_{max}$ to avoid side effects) 
last interactions with the node. If the interaction was a success (the 
transmitted data was the expected data), the mark is $1$, and it is $0$ 
otherwise. The marks included in this list correspond to the own experience 
of node $i$ about node $j$ and are used to compute trust mark for future 
interactions. It is initialized to $\emptyset$.
\medskip

We will call positive (resp. negative) an attribute (valued in $\lbrack 0, 1 
\rbrack$) that is greater or equal to (resp. less than) $\frac{1}{2}$. 

\subsection{Protocol dynamics}

We describe in this subsection the updating methods of the protocol attributes 
described above. First, let us note that the list of external trust knowledges
($et$) and the list of external trust indices in trust knowledges ($iel$) are 
simply transmitted through the network by nodes at each interaction. Then, the 
update of the list of past interactions is done as follows: when node $i$ asks 
node $j$ for data transmission, if the list of past interactions of $i$ is not 
full, the mark affected to this new interaction is placed at the end of the 
list; otherwise, the first mark in the list is removed and the new mark is 
placed at the end. Now, let us focus on other attributes that need more 
complicated updating functions. 
{\vskip 1.5mm}

\noindent {\bf $\bullet$ Trust marks}

When node $i$ wishes to interact with node $j$ at time $t$, it has not only to 
rely on its own knowledges but also on the external knowledges about node $j$ 
it has already received from the other nodes. Thus, to update its trust mark 
about node $j$, node $i$ needs to evaluate two different data. First, we define
the personal knowledges $\mathrm{PK}$ as the simple average of the results of 
past interactions node $i$ had with node $j$. So:

$$
\mathrm{PK_{i,j}^{t+1} = \frac{\sum^{\theta}_{k=1} pil_{i,j}(t)[k]}{\theta}}
$$
{\vskip 1.5mm}

Now, let us also define the external knowledges. Let $\mathrm{S^*}$ be the 
set of nodes $k$ such as $\mathrm{\lbrace k \neq i,j,}$ $\mathrm{iet_{i,k}(t) >
  \frac{1}{2},}$ $\mathrm{et_{i,k}(t)[j] < \frac{1}{2},}$ 
$\mathrm{iel_{i,k}(t)[j] > \frac{1}{2} \rbrace}$. At time $t+1$, the external 
knowledges $\mathrm{EK}$ that node $i$ has about node $j$ are computed as: 

$$
\mathrm{EK_{i,j}^{t+1} = 
  \frac{1}{|S^*|}\sum_{\stackrel{k=1}{\stackrel{S^*}{}}}^{n} et_{i,k}(t) 
  \lbrack j \rbrack}.
$$

The choice made here is to average only the bad opinions other nodes have 
about node $j$. Of course, we will restrict to nodes in which we trust the 
opinion and who themselves trust their own opinion about node $j$. Note that 
this choice leads the protocol to be efficient rather than reliable. 
\smallskip

Now, we update the trust mark by simply taking the weighted sum of these two 
knowledge values:

$$
\mathrm{tm_{i,j}(t+1) = 
\frac{\varepsilon.\lbrack \theta \times PK_{i,j}^{t+1} \rbrack+ 
    \stackrel{(E)}{\overbrace{\zeta.\lbrack \theta_{max} × EK_{i,j}^{t+1}
	\rbrack}}}  {\varepsilon.\theta
  {\underbrace{+ \zeta.\theta_{max}}_{iff~(E) \neq 0}}}}
$$

\noindent where $\varepsilon$ and $\zeta$ are two parameters managing the 
weight of personal and external knowledges. So, the more $\zeta$ has a high 
value compared to $\varepsilon$, the more the diffusion will be fast. However, 
$\varepsilon$ and $\zeta$ have to be well balanced to prevent the fact a node 
is easily disturbed if it has a majority of dishonest neighbours. Let us note 
that, if these two parameters are equal, external knowledges weight is
greater than personal knowledges weight when the $\theta_{max}$ first 
interactions have not already been executed, because $\theta_{max}$ is the 
upper bound of $\theta$. It seems to be reasonable to think that, when a node 
has few personal knowledges about an other with which it wants to interact, 
external knowledges are more important in its decision.

This trust marks evolution method shows that this protocol can not be less 
efficient than another one which does not implement trust diffusion. Indeed, if
an interaction between two nodes $i$ and $j$ is possible, when $i$ updates its 
trust mark about $j$, it can not obtain a higher new mark than one obtained 
only with its past interactions with $j$ since it only uses negative 
external knowledges. 
{\vskip 1.5mm}

\noindent {\bf $\bullet$ Trust indices in personal trust marks}\\
\indent The updating method of the trust mark index (when $i$ asks $j$ for an 
interaction at time $t+1$) is the following. After having 
computed $E(pil_{i,j}(t))$ and $\sigma(pil_{i,j}(t))$ where $E$ and $\sigma$ 
refer respectively to the arithmetic average and the standard deviation, the
new value of the trust mark index is obtained by a threshold algorithm. So, we 
fix a threshold named $\omega_{itm}$ and we verify if the new trust mark
$tm_{i,j}(t+1)$  is in  
$$
\rbrack \mathrm{E(pil_{i,j}(t)) - \omega_{itm}~;~E(pil_{i,j}(t)) + 
  \omega_{itm}} \lbrack
$$ 
\indent The index $itm$ grows if $\alpha$ is in this range of values and 
decreases otherwise. In each case, updates are proportional to 
$\sigma(pil_{i,j}(t))$; however, in order to go one step further in the 
``efficiency'' direction, we have chosen the decreasing function to be faster 
than the increasing one. 
{\vskip 1.5mm}

\noindent {\bf $\bullet$ Trust indices in external trust marks}\\
\indent Node $i$ will use the newly calculated trust mark on node $j$ to adjust
the trust index it has about the external trust knowledges stored for each 
other node. In other words, for each other node $k$, node $i$ compares its new
trust mark about $j$ to the opinion $k$ has about $j$. The more different 
(resp. close) it is, the more node $i$ decreases (resp. increases) its trust 
index on the list of external knowledges given by $k$. A threshold algorithm 
(where the threshold is named $\omega_{iet}$) is used here too. When $i$ 
interacts with $j$, we consider the belonging of each $et_{i,k}(t)[j]$ to the 
following range of values:
$$
\rbrack \mathrm{tm_{i,j}(t+1) ~-~ \omega_{iet} ~;~ tm_{i,j}(t+1) ~+~ 
  \omega_{iet}} \lbrack
$$

If $et_{i,k}(t)[j]$ is in this range of values, the index $iet_{i,k}(t)$ grows,
 otherwise it decreases. There also, the decreasing goes faster than the 
growth.
\medskip

\subsection{Decision to interact}

At time $t+1$, if node $i$ wishes to interact with node $j$, it verifies its 
trust mark about node $j$ and decides to realize the interaction only if this 
trust mark is positive. 

\section{Simulations}
\label{simul}

In this section, we present the simulations that we have done on the trust 
diffusion protocol presented above (called Trudi) and compare the results of 
the latter to the so-called {\em $\beta$-protocol} in which nodes do not 
transmit their knowledges about other nodes. We are going to evaluate the 
efficiency of the protocols by measuring three parameters: (i) the number of 
honest nodes succeeding in identifying the dishonest nodes; (ii) the total 
number of bad interactions (transmitted data are not the expected ones) that 
the dishonest nodes succeeded to do; (iii) in some cases, we also measure the
number of interactions needed for the honest nodes to identify the dishonest 
ones. 

Let us precise the meaning of {\em honest} and {\em dishonest} we will use in 
the following.  We will call {\em action-dishonest} (resp. {\em action-honest})
a node that wishes to give to every other unexpected data (resp. expected data)
and {\em opinion-dishonest} (resp. {\em opinion-honest}) a node that wishes to 
give wrong opinions (resp. correct opinions) about some nodes to many others. 
A node is called {\em dishonest} if it is either action-dishonest or 
opinion-dishonest and {\em totally-dishonest} if it is both action-dishonest 
and opinion-dishonest.

Moreover, a trust protocol is called {\em coherent} if, when a little 
proportion of the system actors is action-dishonest, the honest ones succeed in
the detection of those latter. 
\medskip

We will start by describing the simulation protocol used in the study before 
exposing some of the obtained results. 

\subsection{Simulation protocol}

In order to run simulations not to far from real interactions in ad hoc 
networks, we are going to make the hypothesis that if we observe the set of 
interactions in the network after a sufficient amount of time, we will observe 
a network close either to a social network or to a complete network (where each
node has interacted at least once with any other one). A social network is the 
personal or professional set of relationships between individuals. It has been 
shown~\cite{albar,newman} that these networks very often present the following 
characteristics: their degree distribution follows a power law (the probability
that a node has a degree $k$ is a decreasing power of $k$); they present a 
clustering coefficient (the probability for two nodes having a common neighbour
to be connected) significantly higher than for classical random 
networks~\cite{strogatz_watts}. Of course, there is no evidence that these 
social networks precisely corresponds to the reality, but it makes sense to 
suppose that interactions in ad hoc networks will share some features with 
interactions in social networks. 

In order to run simulations, that will lead at the end to such interaction 
networks, we determine the interaction graph {\it a priori} and we then realize
the interactions on this graph. Thanks to this method, we control the final 
shape of the interactions superposition. Because of the above hypothesis, the 
chosen graphs are either power law degree distribution graphs with high 
clustering coefficient (such a network will be called ``social network'' in the
following) or complete graphs. 

\subsection{Simulations results}

The results presented in this subsection come from simulations of $1000000$ 
interactions on $100$ nodes networks.

In the first simulation, there is only one action-dishonest node which sends 
bad data with probability $\frac{1}{5}$ and no opinion-dishonest node. In the 
social network, this node is given the highest degree. We measure the average 
and maximum number of interactions realized by the other nodes with this 
action-dishonest one. In Table~\ref{tab_simple}, which presents the results of 
this simulation, we observe that Trudi is significantly better than the 
$\beta$-protocol. Indeed, the perception of dishonesty is about $45$ times 
faster with Trudi than with the $\beta$-protocol. This result can be explained 
by the nature of trust diffusion in Trudi. Only negative marks and opinions are
used to realize the updates of the protocol attributes. Therefore, as the 
weight of external knowledges is always more important than this of personal 
knowledges, honest nodes perceive quasi-directly the dishonesty. 

Another simulation has been run by placing the dishonest node on the lowest 
degree node. In this case, there is no real difference between the two 
protocols because of the too small connectivity of the node. Thus, the interest
of the protocol depends on the probability dishonest nodes have to give bad 
interactions and on the place of dishonesty in the network, an effect that we 
will call {\em topology dependence}. 
\begin{center}
  \begin{table}[h]
    \caption{Comparison of Trudi and the $\beta$-protocol after a simple 
      attack.}
    \begin{center}
      \begin{tabular}{|c|c|c|} \hline
	& Trudi & $\beta$-protocol \\ \hline\hline
	Avg ; Max & $3.38~;~10$ & $153.04~;~\sim 420$ \\ \hline
      \end{tabular}
    \end{center}
    \label{tab_simple}
  \end{table}
\end{center}
Let us now assume that some nodes make a coalition in order to attack the 
network, {\it i.e.} to use their knowledge about the protocol to break its 
efficiency or reliability. We are going to present in the following some 
results describing the effects that such an attack can have, with and without 
the Trudi protocol.  
{\vskip 1.5mm}

\noindent {\bf $\bullet$ Simple coalition attacks}\\
\indent {\em A simple coalition attack manages a group of nodes in which the 
  members are honest with each other but totally-dishonest with nodes that are 
  not a part of the group, {\it i.e.} they protect themselves by giving 
  positive opinions about the group members outside the group.}
{\vskip 1.25mm}

We have run simulations of this type of attack on complete networks. In each 
simulation, we have increased the size of the group of dishonest nodes (which 
send bad data with probability $\frac{1}{5}$) to focus on the minimal size this
group must have to destabilise the network, {\it i.e.} at least one honest node
does not succeed in perceiving all the dishonest ones. 
\begin{center}
  \begin{table}[h]
    \caption{Comparison of Trudi and the $\beta$-protocol after coalition 
      attacks.}
    \begin{center}
      \begin{tabular}{|c|c|c|} \hline
	Number of dishonest nodes in the group & Trudi & $\beta$-protocol \\ 
	\hline\hline
	40 & $\surd$ & $\times$ \\ \hline
	50 & $\surd$ & $\times$ \\ \hline
	60 & $\surd$ & $\times$ \\ \hline
	70 & $\times$ & $\times$ \\ \hline
	80 & $\times$ & $\times$ \\ \hline
      \end{tabular}
    \end{center}
    \label{tab_coalition}
  \end{table}
\end{center}
Table~\ref{tab_coalition} shows that Trudi is extremely robust against simple 
coalitions contrary to the $\beta$-protocol. With Trudi, the system can 
actually be destabilised only if the rate of good interactions is much less 
than the rate of bad interactions provided by dishonest nodes. Let us 
illustrate these results on a simple example which models a complete network 
where every dishonest node has a probability of $\frac{1}{5}$ to provide a bad 
interaction. At each interaction, an edge has a probability of $\frac{1}{9900}$
to be chosen (the interactions from $i$ to $j$ 
and from $j$ to $i$ are differentiated). So, when only thirty nodes are honest,
the probability for two honest nodes to interact is $\frac{30 \times 29}{9900} 
\approx 0.0878$. When forty nodes are honest, this probability is $\frac{40 
\times 39}{9900} \approx 0.1575$. So, in the second case, the system is not 
destabilised because the rate of good knowledges diffusion is close to the rate
of bad interactions provided by dishonest nodes. This explanation has been 
verified by other simulations.
{\vskip 1.25mm}

Moreover, note that the results  presented in Table~\ref{tab_coalition} are 
also valid for social networks despite the topology-dependence.
{\vskip 1.25mm}

Let us now considerer more sophisticated attacks.
{\vskip 1.50mm}

\noindent {\bf $\bullet$ Trojan attacks}\\
\indent {\em A Trojan attack (by analogy to the mythical war of Troy) manages 
  a group of nodes where one of them (Ulysses) is honest inside the group and 
  totally-dishonest outside the group. Ulysses is protected by the other 
  members of the group, {\it i.e.} they are action-honest but they always 
  provide positive opinion to the honest nodes about Ulysses.}
{\vskip 1.25mm}

For this attack, two simulations have been executed (illustrating protocols
robustness both on complete and social networks) to see the number of bad 
interactions ($nbi$) realized by Ulysses (which sends bad data with probability
$\frac{1}{5}$) and the maximum number of interactions ($mni$) executed by a 
honest node with Ulysses before perceiving its dishonesty. 
\begin{center}
  \begin{table}[h]
    \caption{Comparison of Trudi and the $\beta$-protocol after Trojan 
      attacks.}
    \begin{center}
      \begin{tabular}{|c|c|c|} \hline
	Type of network and & Trudi & $\beta$-protocol \\
        Ulysses error probability & & \\\hline\hline
	Complete network - $20\%$ & $\mathrm{nbi=52}$ ; $\mathrm{mni=8}$ & 
	$\mathrm{nbi=1233}$ ; $\mathrm{mni=130}$ \\ 
	\hline
	Social network - $20\%$ & $\mathrm{nbi=14}$ ; $\mathrm{mni=25}$ & 
	$\mathrm{nbi=344}$ ; $\mathrm{mni \sim 900}$ \\ \hline
      \end{tabular}
    \end{center}
    \label{tab_trojan}
  \end{table}
\end{center}
In the second simulation, Ulysses is the highest degree node and is protected 
by $9$ high degree nodes. Table~\ref{tab_trojan} validates the robustness of 
Trudi against Trojan attacks on complete networks as well as on social 
networks. However, other simulations where the dishonest group takes place on 
nodes of lowest degrees have shown that Trudi efficiency is not really 
significantly more interesting than one of the $\beta$-protocol. These results 
confirm that topology dependence is the main factor of the interest of the 
protocol in social networks.
{\vskip 1.5mm}

\noindent {\bf $\bullet$ Detonator attacks}\\
\indent {\em A detonator attack is a Trojan attack where the dishonesty of 
  Ulysses is started only after a determined time which is called the 
  detonation time.}
{\vskip 1.25mm}

Our objective is to determine how many bad interactions ($nbi$) are provided by
Ulysses, after its detonation, to be recognised as a mistrustable node and 
how many honest nodes ($nhn$) succeed in perceiving the dishonesty of Ulysses. 
\begin{center}
  \begin{table}[h]
    \caption{Comparison of Trudi and the $\beta$-protocol after detonator 
      attacks.}
    \begin{center}
      \begin{tabular}{|c|c|c|} \hline
	Number of interaction & Trudi & $\beta$-protocol \\ 
        after the detonation & & \\ \hline\hline
	Complete network & $\mathrm{nbi=836}$ ; $\mathrm{nhn=19}$ & 
	$\mathrm{nbi=837}$ ; $\mathrm{nhn=12}$ \\ 
        500000 & & \\\hline
	Complete network & $\mathrm{nbi=1051}$ ; $\mathrm{nhn=75}$ & 
	$\mathrm{nbi=1411}$ ; $\mathrm{nhn=27}$ \\ 
        900000 & & \\ \hline\hline
	Social network & $\mathrm{nbi=419}$ ; $\mathrm{nhn=15}$ & 
	$\mathrm{nbi=657}$ ; $\mathrm{nhn=15}$ \\ 
        900000 & & \\ \hline
      \end{tabular}
    \end{center}
    \label{tab_detonator}
  \end{table}
\end{center}
Two simulations have been run to show the robustness of the two protocols 
against detonator attacks in complete networks where Ulysses sends bad data 
with probability $\frac{1}{5}$ and is protected by $9$ other nodes.

The first (resp. the second) simulation illustrates this type of attack where 
Ulysses becomes dishonest after the $500000$th (resp. $100000$th) 
interaction. Table~\ref{tab_detonator} shows that a too little number of 
interactions executed after the detonation of Ulysses node prevents many honest
ones to perceive its dishonesty. Indeed, as the interactions are chosen 
randomly in the $n \dot (n-1)$ possible, in a $100$ nodes system, $9900$ 
($\approx 10000$) different interactions exist. 
Consequently, an edge between a given node and Ulysses is potentially chosen
fifty times in the first simulation. It is not sufficient for the given node to
detect the dishonesty of Ulysses because of two reasons. Firstly, the given 
node has already a positive opinion about Ulysses and, secondly, the latter 
has a too small probability to provide bad interactions ($p = \frac{1}{5}$). 
The second line of Table~\ref{tab_detonator} validates this hypothesis. If we 
focus on the number of bad interactions, we can conclude our protocol is 
interesting because it provides a mean of $11.68$ of bad interactions to each 
honest node. 

Then, another simulation with the same features than the second one has been 
run to show the robustness of the two protocols against these attacks in social
networks. Ulysses is here the node with the highest degree (with $15$ honest
neighbours).  The last line of Table~\ref{tab_detonator} shows that, with the 
two protocols, all Ulysses neighbours succeed in perceiving its dishonnesty. 
However, with Trudi, Ulysses node gives less bad data than with the 
$\beta$-protocol. Moreover, if we focus on the total number of interactions 
realized by the $15$ neighbours of Ulysses after the detonation time, we see 
that this number is approximately three times more important with the 
$\beta$-protocol than with Trudi.  

\section{Conclusion and perspectives}
\label{perspec}

The results presented in this paper show that such a trust diffusion protocol 
could have interesting efficiency properties. Besides, it could apply not only 
in ad hoc networks but also in other networks. It could be of interest in 
Internet and P2P systems. For example, it could be useful to measure trust 
users give to web sites and could broadcast sites dishonesty. However, to 
obtain such protocols, improvements have to be done. 
\smallskip

In particular, future research should focus on the reduction of the size of 
the transmitted messages because the protocols ``lightness'' is a primordial 
point to avoid the network surcharge. So, we should try to find a good balance 
between reducing the transmitted messages size and the interest of the 
diffusion principle. Different solutions could be imagined. For example, since 
the goal of the protocol is to ensure efficiency, we could only diffuse the 
information about bad behaving nodes. Another solution would be to diffuse only
a proportion (function of the network size) of these negative knowledges. Such 
hypotheses should be tested via new simulations processes to be validated. 

{ 
  \bibliographystyle{alpha}
  \bibliography{ms06}
}

\end{document}